\documentclass[twocolumn,pre]{revtex4}
\usepackage{amssymb,hyperref}
\usepackage{graphicx}
\usepackage{subfigure}

\begin{document}

\title{Competition between Multiple Totally Asymmetric Simple Exclusion Processes for a
Finite Pool of Resources}
\author{L. Jonathan Cook}
\email{lacook1@vt.edu}
\author{R. K. P. Zia}
\email{rkpzia@vt.edu}
\author{B. Schmittmann}
\email{schmittm@vt.edu}
\affiliation{Department of Physics, Virginia Tech, Blacksburg, VA 24061, USA}
\date{\today }

\begin{abstract}
Using Monte Carlo simulations and a domain wall theory, we discuss the effect of
coupling several totally asymmetric simple exclusion processes (TASEPs) to a 
finite reservoir of particles. This simple model mimics directed biological 
transport processes in the presence of finite resources, such as protein 
synthesis limited by a finite pool of ribosomes. If all TASEPs have equal
length, we find behavior which is analogous to a single TASEP coupled to a 
finite pool. For the more generic case of chains with different lengths, several
unanticipated new regimes emerge. A generalized domain wall theory captures
our findings in good agreement with simulation results.   
\end{abstract}
\maketitle



\section{Introduction}

A fundamental and comprehensive understanding of non-equilibrium phenomena
remains one of the greatest challenges of current condensed matter and
materials physics \cite{CMMP2010}, with significant consequences for
advances in materials science, the life sciences, and engineering. Even the
non-equilibrium steady states of open, transport-carrying systems continue
to defy our equilibrium-trained expectations and intuitions. One approach
towards progress has focused on investigating simple model systems, with the
goal of identifying generic classes of behaviors.

The totally asymmetric simple exclusion process (TASEP) \cite
{Derrida98,Schutz01} is one of these models. It has acquired paradigmatic
status for several reasons:\ (i) it is very simple, consisting of particles
hopping along a one-dimensional chain; (ii) with open boundaries, it shows
highly nontrivial behaviors, such as distinct phases, shocks, and long-range
correlations in both space and time; (iii) in its simplest forms, its
steady-state properties, as well as selected dynamic quantities, can be
found exactly; and finally, (iv) the model is closely related to interesting
applications, such as biological transport \cite{bio} or traffic flow \cite
{Traffic}. At the origin of this richness lies the violation of detailed
balance. The specific behaviors depend strongly on the boundary conditions.
With periodic boundary conditions, the stationary state is a flat
distribution with all configurations equally probable \cite{Spitzer70}.
However, the dynamics of this system is nontrivial and differs from simple
diffusion \cite{RingDyn}. With open boundary conditions, particles are
injected at one end and removed at the other with different (but constant)
rates. In this case, even the steady state is nontrivial and remained
unknown for two decades \cite{Derrida92,Derrida93,Schutz93}. Despite being a
one-dimensional system with short range interactions and dynamics, the open
TASEP displays distinct (stationary) phases \cite{Krug91}, controlled by the
entrance and exit rates. As may be expected, the dynamic properties are even
more complex and rich \cite{OpenDyn,DW,Santen02,AZS07}.

These TASEP studies have recently been extended by coupling the chain to a 
\emph{finite} (rather than infinite)\ particle reservoir \cite
{HaDenN02,ASZ08}, reflecting a constraint on the total number of particles.
In \cite{HaDenN02}, the particles represent cars leaving a parking garage,
so that the rate of entry onto the roadway (the lattice) is chosen to be a
constant, as long as there is at least one car in the garage. In \cite{ASZ08}%
, the TASEP models a biological transport process \cite{bio}, and the
constraint reflects the finite number of ribosomes in a cell, with those
leaving the mRNA (the lattice) being ``recycled'' to the entry point. The
origins and effects of ``ribosome recycling'' are multifaceted, such as
diffusion of the recycled components from termination to initiation sites
\cite{TChou03}. Addressing all possible issues for a real cell is beyond the
scope of this study and our aim here is modest, namely, to explore how the finite pool
of available particles affects the lattice occupation and currents. As in 
\cite{ASZ08}, we consider an entry rate that depends smoothly on the number
of particles in the pool, $N_p$. In particular, we choose a rate function
which is proportional to $N_p$ when the concentrations of ribosomes in the
cell are limited, and, when particles are abundant, becomes a constant --
the inherent binding rate of a ribosome. The simulation results for 
a \emph{single} TASEP can be described well by a domain wall (DW)
theory \cite{DW,Santen02,ASZ08}, especially when generalized to account for
the feedback \cite{CZ08}.

In addition to being constrained to a finite pool of ribosomes in a living
cell, a mRNA must compete against many other different genes (or mRNAs) for
these resources. Therefore, we are motivated to study the competition of
multiple TASEPs for the same pool of particles. Since different mRNAs are
comprised of different numbers of codons, it is reasonable to study TASEPs
with different lengths. Does one ``win'' and another one ``lose''? What does
``winning'' and ``losing'' mean?

This paper is organized as follows. The next section introduces our model.
In Section \ref{sect3}, we present simulation results for two and three
TASEPs connected to a finite reservoir of particles. Analytic results, based
on our generalized DW theory \cite{CZ08}, are derived for an arbitrary
collection of TASEPs and will be discussed in Section \ref{sect4}. A final
section is devoted to a summary and outlook.

\section{A model for competing TASEPs}

Let us first review the standard (or ``ordinary'')\ TASEP. Each site of a
discrete lattice of length $L$, is labeled by $i=1,...,L$ and may be empty
or occupied by a single particle. Thus, a particular configuration can be
described by a set of occupation numbers, $\{n\left( i\right) \}$, each
taking the values $0,1$. A configuration evolves to a new one according to
the following rules. A particle is chosen at random, say, at site $i$. If
the neighboring site to its right (site $i+1$) is empty, the particle hops
there with rate unity. If the particle is located on the last site ($i=L$)
of the lattice, it exits with a probability $\beta $. In addition to the
particles on the lattice, we assign a ``virtual\ particle'' to label an
external reservoir, so that when chosen, a particle is placed with
probability $\alpha $, on the first site ($i=1$) of the lattice, provided $%
n\left( 1\right) =0$. Notice that this system can be regarded as a total of $%
N_{tot}$ particles hopping on a ring with $L+1$ sites, where the site $i=0$
is associated with the reservoir and can be occupied by any number of
particles. Of course, the hopping rates into and out of this site are
different from the rest. They can be represented, respectively, as $\beta
n\left( L\right) $ and $\alpha \left[ 1-n\left( 1\right) \right] $. Note
that neither depends on $n\left( 0\right) $ and so, as long as $N_{tot}>L$ ,
there would be at least one particle which can be injected onto the lattice.
Indeed, this is the scenario presented in the parking garage problem \cite
{HaDenN02}, in which interesting transitions are studied for $N_{tot}\in
\left[ 0,L\right] $. Lastly, the seemingly complicated rule of choosing real
and virtual particles can be replaced, in this formulation of TASEP, by:
``Randomly choose an occupied site.''

In the steady state, this simple model exhibits three phases:\ a low density
(LD) phase for $\alpha <1/2$ and $\alpha <\beta $, a high density (HD) phase
for $\beta <1/2$ and $\beta <\alpha $, and a maximal current (MC) phase for $%
\alpha ,\beta >1/2$. The densities $\bar{\rho}$ in the three phases are
given by $\alpha $ (LD), $1-\beta $ (HD), and $1/2$ (MC), respectively. The
line $\alpha =\beta <1/2$ marks the coexistence of HD and LD phases and is
characterized by the presence of a shock which performs an unbiased random
walk through the whole lattice. For this reason, the coexistence line is
also sometimes referred to as the ``shock phase'' (SP). Since the entrance
and exit rates are constant here, $N_{tot}$ plays no role and can be finite
(but larger than $L$) instead of $\infty $.

Our motivation for studying the TASEP is modeling protein synthesis in a cell
\cite{bio}, in which the particles represent ribosomes. Thus, $N_{tot}$ is
finite and must be shared between many mRNA's (there may be as many as 
10,000 mRNAs in some cells). Only the unbound ribosomes (i.e., those in the
``pool'', totalling $N_p$) are available for initiation (i.e., to enter any
one TASEP). Assuming the concentration of these ribosomes is uniform, we
introduced a more realistic model \cite{ASZ08} for the entry rate, $\alpha $%
, that depends on $N_p$ as follows. Starting with just one lattice (mRNA) in
our system, let us denote its total occupation by $N$, so that 
\begin{equation}
N_{tot}=N+N_p  \label{Ntot=}
\end{equation}
is kept fixed. Particles still exit the lattice as before, with rate $\beta $%
, and are placed immediately into the pool. In contrast to the ordinary TASEP,
the (effective) entry rate, $\alpha _{eff}$, will now be controlled by $N_p$%
, through $\alpha _{eff}=\alpha f(N_p)$. Here, $\alpha $ is a constant and $%
f(N_p)$ assumes a value in $\left[ 0,1\right] $ for all $N_p$. Physically, $%
\alpha _{eff}$ should vanish if there are no particles in the pool, whence
we impose $f(0)=0$. Further, $f$ should be a monotonically increasing
smooth function of $N_p$. Finally, for sufficiently large $N_p$, we should
recover the standard TASEP with a constant entry rate, $\alpha $, whence 
$\lim_{N_p\rightarrow \infty }f(N_p)=1$. These properties are motivated
by the notion that the rate for a ribosome to bind to the mRNA (a) should be
limited by the ribosome availability, especially for low concentrations and
(b) should approach some intrinsic rate for the binding process, when the
ribosomes are abundant. The specific form of $f(N_p)$ is not very important
but we follow \cite{ASZ08} and use 
\begin{equation}
f(N_p)=\tanh (N_p/N^{*})  \label{f(Np)}
\end{equation}
where $N^{*}$ is a suitably defined normalization, or crossover, scale. In 
\cite{ASZ08}, $N^{*}$ was chosen to be the average number of particles for
the standard TASEP\ with entry and exit rates $\alpha $\ and $\beta $.

In the following, we will consider multiple open TASEPs of various lengths, $%
L_\ell $ ($\ell =1,2,...M$), $M$ being the total number of chains in our
system. Writing the occupation in each as $N_\ell $, we naturally define 
\begin{equation}
\rho _\ell \equiv N_\ell /L_\ell 
\end{equation}
as the overall densities on each chain and of course, 
\begin{equation}
N_{tot}=\sum_{\ell =1}^MN_\ell +N_p  \label{Ntot-def}
\end{equation}
as the generalization of Eqn. (\ref{Ntot=}). Most of our simulation results are for $M=2$, with a few for $M=3$. All $M$ TASEPs draw their 
particles from the \emph{same} reservoir, according to the \emph{same} 
$\alpha _{eff}=\alpha f(N_p)$ with $f$ given by Eqn. (\ref{f(Np)}). 
For multiple chains of different lengths, we choose $N^{*}$ to be 
\begin{equation}
N^{*}=M^{-1}\sum_{\ell =1}^M\bar{\rho}_\ell L_\ell   \label{N_star}
\end{equation}
where $\bar{\rho}_\ell $ is the average density for an ordinary TASEP of
length $L_\ell $, with entry and exit rates $\alpha $\ and $\beta $. This
choice of $N^{*}$ only serves to locate the specific value of $N_{tot}$ at
which the system crosses a phase boundary. It also allows us to define
(somewhat arbitrarily) 
\begin{equation}
\rho_{tot}\equiv N_{tot}/M^{-1}\sum_{\ell=1}^ML_\ell\,\,.
\end{equation}

For our simulation studies, we define a Monte Carlo Step (MCS) as making $M + \sum L_\ell$ attempts to 
choose a site to update. Note that, on the average, the pool is updated $M$ times as often as a site on any lattice. This choice corresponds to updating all links (nearest neighbor pairs of sites) with equal probability, with the understanding that each TASEP is linked to the pool (independently of the others). In this study, the entry rates to all chains are chosen to be the same $\alpha_{eff}$. We typically start with all particles in the pool and wait 100k MCS for the 
system to reach steady state. Every 100 MCS thereafter, we measure the site occupations, $n\left( i\right) $, in each chain. Using up to 10k measurements
as our ``ensemble,'' we compute the density profile 
\begin{equation}
\rho \left( i\right) \equiv \left\langle n\left( i\right) \right\rangle 
\end{equation}
and the overall density $\rho \equiv \sum_i\rho \left( i\right) /L$ for 
each chain. The average steady-state current, denoted by $J$, is obtained by
monitoring the total number of particles that enter (and exit) a chain over
the last 1M MCS. Note that $J$ is also just the product $\beta \rho
\left( L\right) $.

\section{Simulation Results}

\label{sect3}

In this section, we present simulation data for the model described above.
In principle, any number of TASEPs can be connected to the pool. However, to
gain some insight into competition, we begin with the simplest case, with
only two TASEPs. Although some selected results for the $M=3$ case will be
presented at the end, most of our data here is for the case of $M=2$. As we
will see, novel features already appear when just one more TASEP is
added to the system. By contrast, we have not encountered, so far, any
further surprising phenomena by considering $M>2$. As in the earlier study 
\cite{ASZ08}, we explore four regimes here, with $\alpha $ and $\beta $
corresponding to the LD, HD, MC phases of the ordinary TASEP, as well as the
coexistence line, SP. Our main interest is how an increasing $N_{tot}$
affect the various profiles and so, the average overall densities and
currents.

\subsection{LD phase}

\begin{figure}
\subfigure[]{\includegraphics[width=8.2cm]{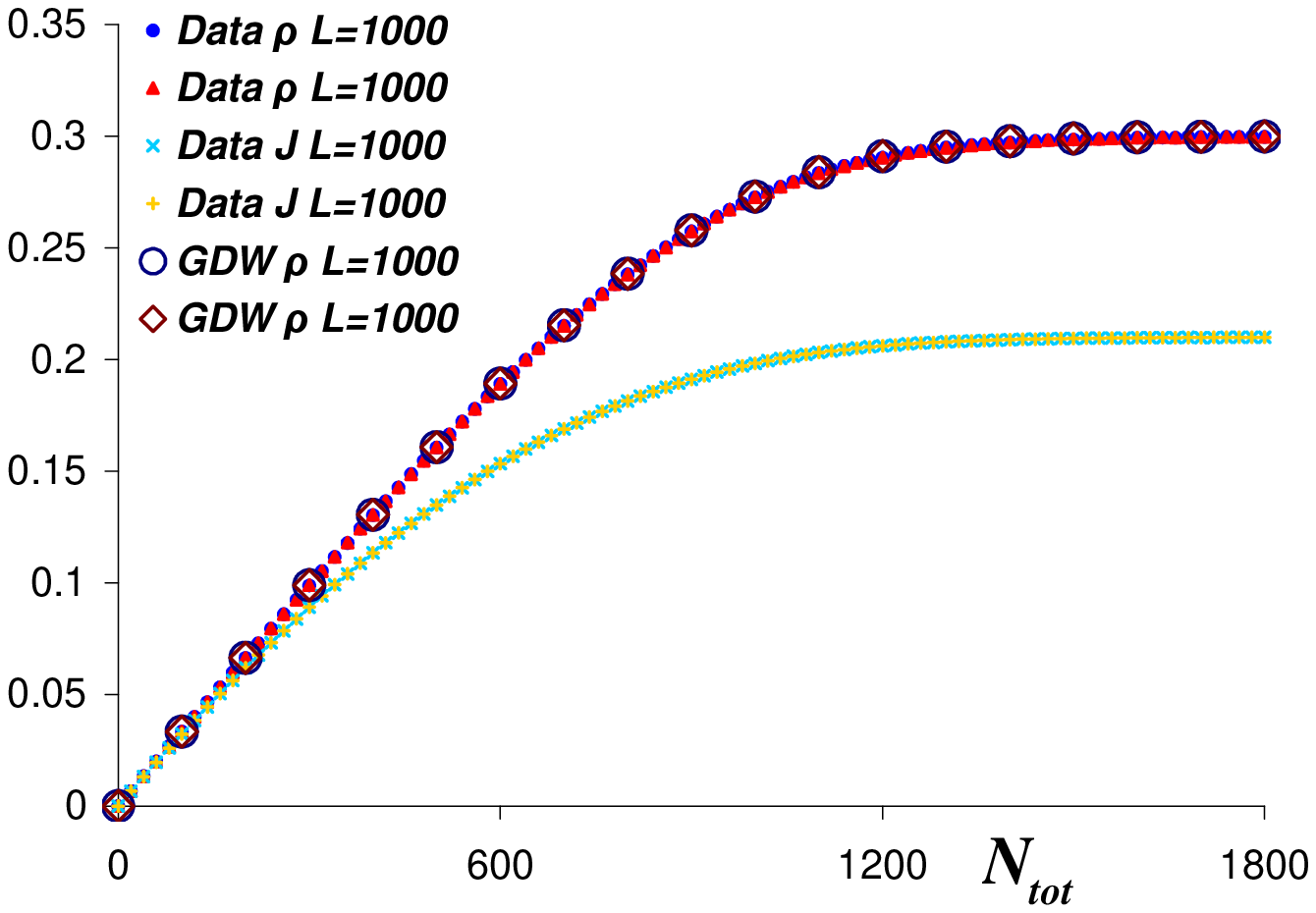} \label{LD-density-equal} }
\subfigure[]{\includegraphics[width=8.2cm]{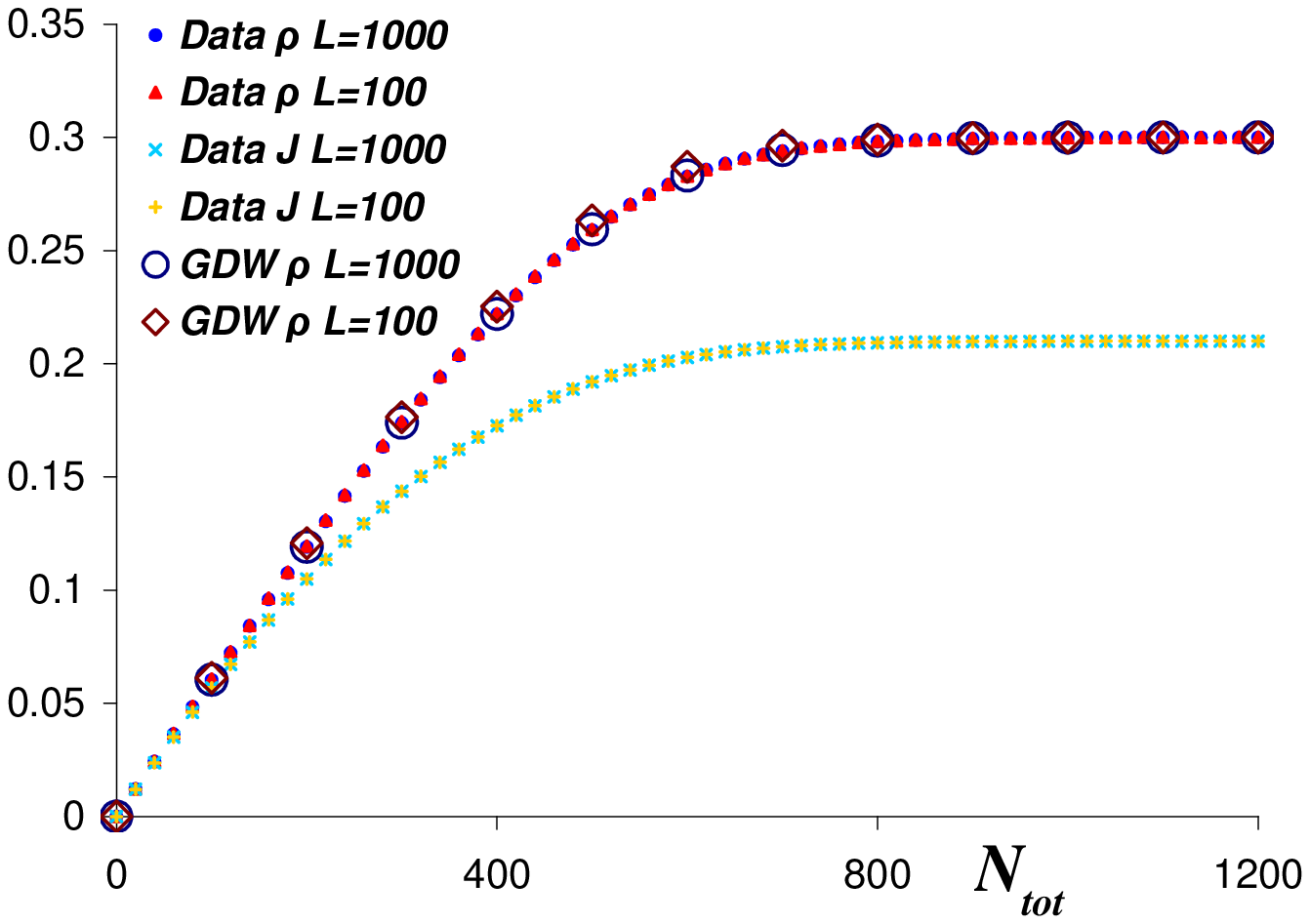} \label{LD-density-unequal} }
\caption{(Color online) Average overall density and current as a function of $N_{tot}$ for \subref{LD-density-equal} 
$L_1=L_2=1000$ and \subref{LD-density-unequal} $%
L_1=1000$ and $L_2=100$ with $\alpha =0.3$ and $\beta =0.7$.\label{LD-density}}
\end{figure}

When $\alpha <1/2$ and $\alpha <\beta $, the standard TASEP displays the LD
phase. If coupled to a finite pool, it remains in the LD\ phase, since $%
\alpha _{eff}$ $\leq \alpha $. The only difference is that the density and
current of the system rise linearly with $N_{tot}$ for small values of $%
N_{tot}$, and approach their asymptotic values from below as $%
N_{tot}\rightarrow \infty $. When two TASEPs compete for this finite pool of
resources, similar behavior is found. The results for equal length chains
are illustrated in Fig. \ref{LD-density-equal} for $\alpha =0.3$, $\beta
=0.7 $, and $L_1=L_2=1000$. As $N_{tot}$ is increased, the two TASEPs both
remain in the LD phase, with equal densities, $\rho _1=\rho _2=\rho $, and
currents, $J_1=J_2=J$. Since both TASEPs are controlled by the same $\alpha
_{eff}$ and $\beta $, there is full symmetry between the two and the
observed behavior is hardly remarkable. The only notable difference between
our system and the one with a single constrained TASEP is observed in 
$\rho \left( N_{tot}\right) $ for $N_{tot}\ll \alpha L$. For our case, 
$\rho \rightarrow \rho _{tot}/3$ as opposed to $\rho _{tot}/2$ for the single
TASEP. This property is easily
understandable, since we can regard the pool as an additional ``chain'' and note 
that the resources are evenly distributed amongst three (or two) ``chains.'' 
For large $N_{tot}$,
both TASEP densities level off to $\alpha $, of course. Slightly more
interesting is the case where the two chains have very different lengths,
such as $L_1=1000,$ $L_2=100$. Due to the lack of symmetry, it is less obvious
that $\rho _1=\rho _2\rightarrow \rho _{tot}/3$ is still valid for small $%
\rho _{tot}\ll \alpha $. However, it is straightforward to show, using the
methods in previous studies \cite{ASZ08,CZ08}, that $\rho _1=...=\rho
_M\rightarrow \rho _{tot}/\left( M+1\right) $ in general, given the specific
form of $\alpha _{eff}\left( N_p\right) $ we chose. At the opposite limit,
the approach to the asymptotic regime is somewhat faster than for the equal
length case. This behavior is also expected, since the ``total'' system size
($L_1+L_2=1100$) is considerably smaller than the case above ($L_1+L_2=2000$%
), so that saturation sets in at smaller values of $N_{tot}$. 

To summarize, the overall densities of competing chains behave just as a single TASEP coupled to a finite pool of particles. Meanwhile, the overall currents are, except for finite size effects, also essentially the same: 
$\rho ( 1 - \rho )$.

\subsection{MC phase}

\begin{figure}
\subfigure[]{\includegraphics[width=8.2cm]{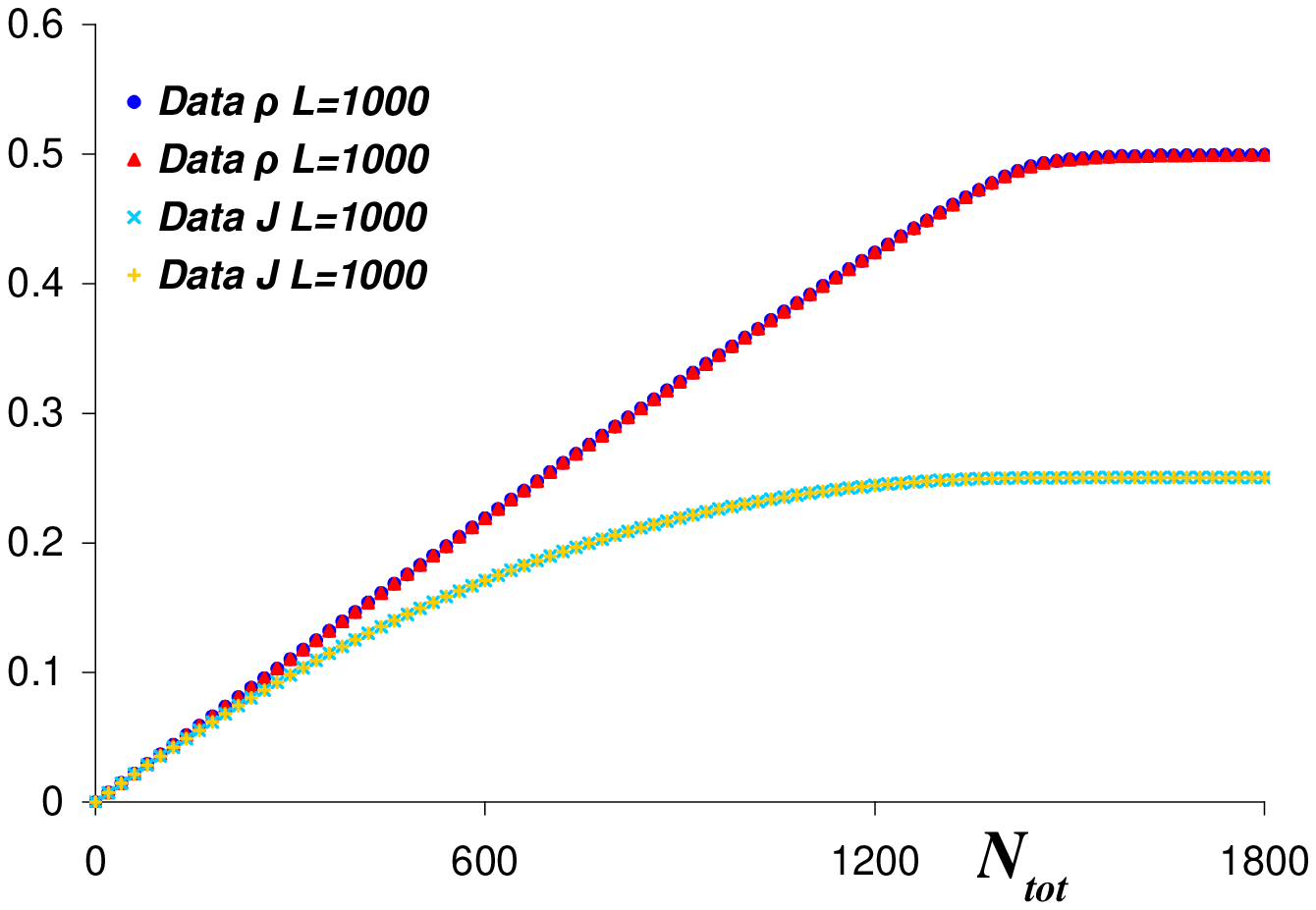} \label{MC-density-equal} }
\subfigure[]{\includegraphics[width=8.2cm]{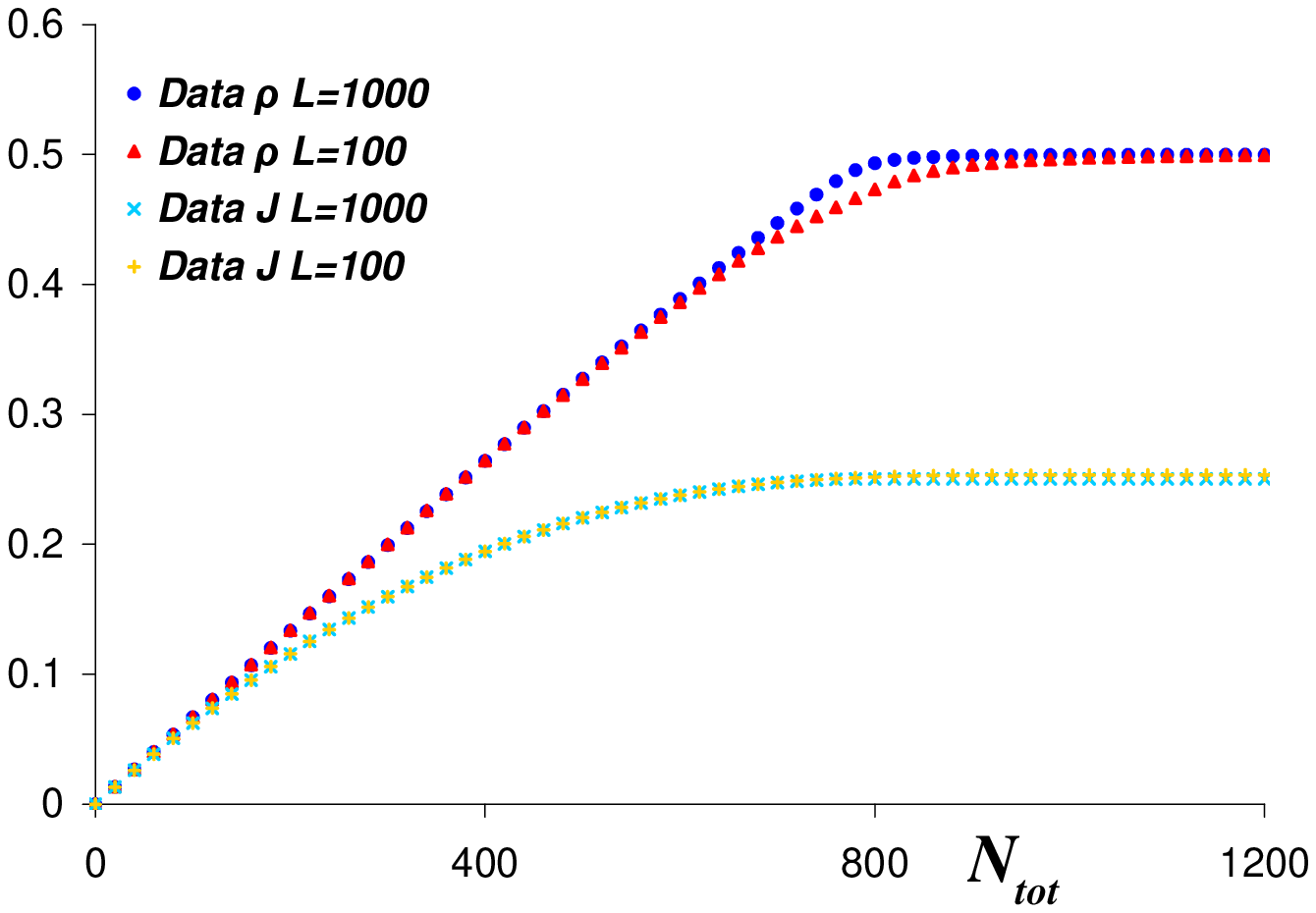} \label{MC-density-unequal} }
\caption{(Color online) Average overall density and current as a function of $N_{tot}$ for 
\subref{MC-density-equal} $L_1=L_2=1000$ and \subref{MC-density-unequal} $%
L_1=1000$ and $L_2=100$ with $\alpha =\beta =0.7$.\label{MC-density}}
\end{figure}

For a single TASEP coupled to a finite pool with $\alpha ,\beta >1/2$, the
current approaches its asymptotic value ($1/4$) smoothly from below.
However, the density exhibits a sharp ``kink,'' which marks the crossing of
the LD-MC phase boundary when $N_{tot}$ becomes large enough to sustain a
density of $1/2$ on the chain. The range of $N_{tot}$ over which this
crossover occurs is controlled by the system size: For large system sizes,
it becomes very narrow and therefore, appears as a ``kink''; in smaller
systems, the crossover occurs over a larger range and appears smooth. For
the case of two TASEPs, we observe very similar behavior. Fig. \ref
{MC-density} shows our data for $\alpha =\beta =0.7$ and $L_1=L_2=1000$.
Again, the currents and densities of the two TASEPs coincide, provided the
two TASEPs have equal lengths. For unequal lengths, we observe the
anticipated finite-size effect in the density:\ The longer TASEP displays a
much sharper crossover from LD to MC behavior than the shorter one.  Like the LD case, the overall densities and currents of two competing chains behave much like those for a single TASEP, including the finite size effects. 

\subsection{SP}

\begin{figure}
\subfigure[]{\includegraphics[width=8.2cm]{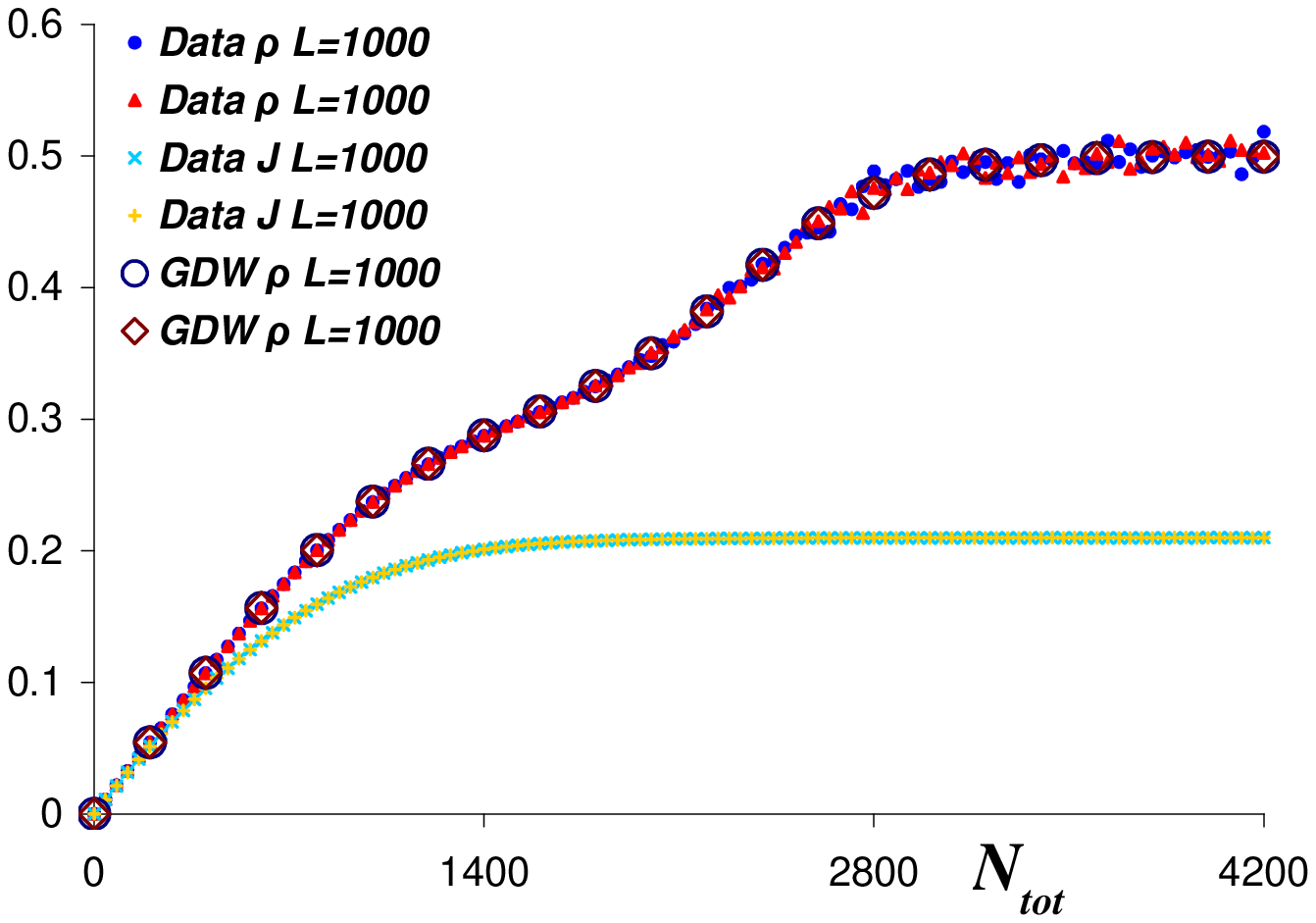} \label{SP-density-equal} }
\subfigure[]{\includegraphics[width=8.2cm]{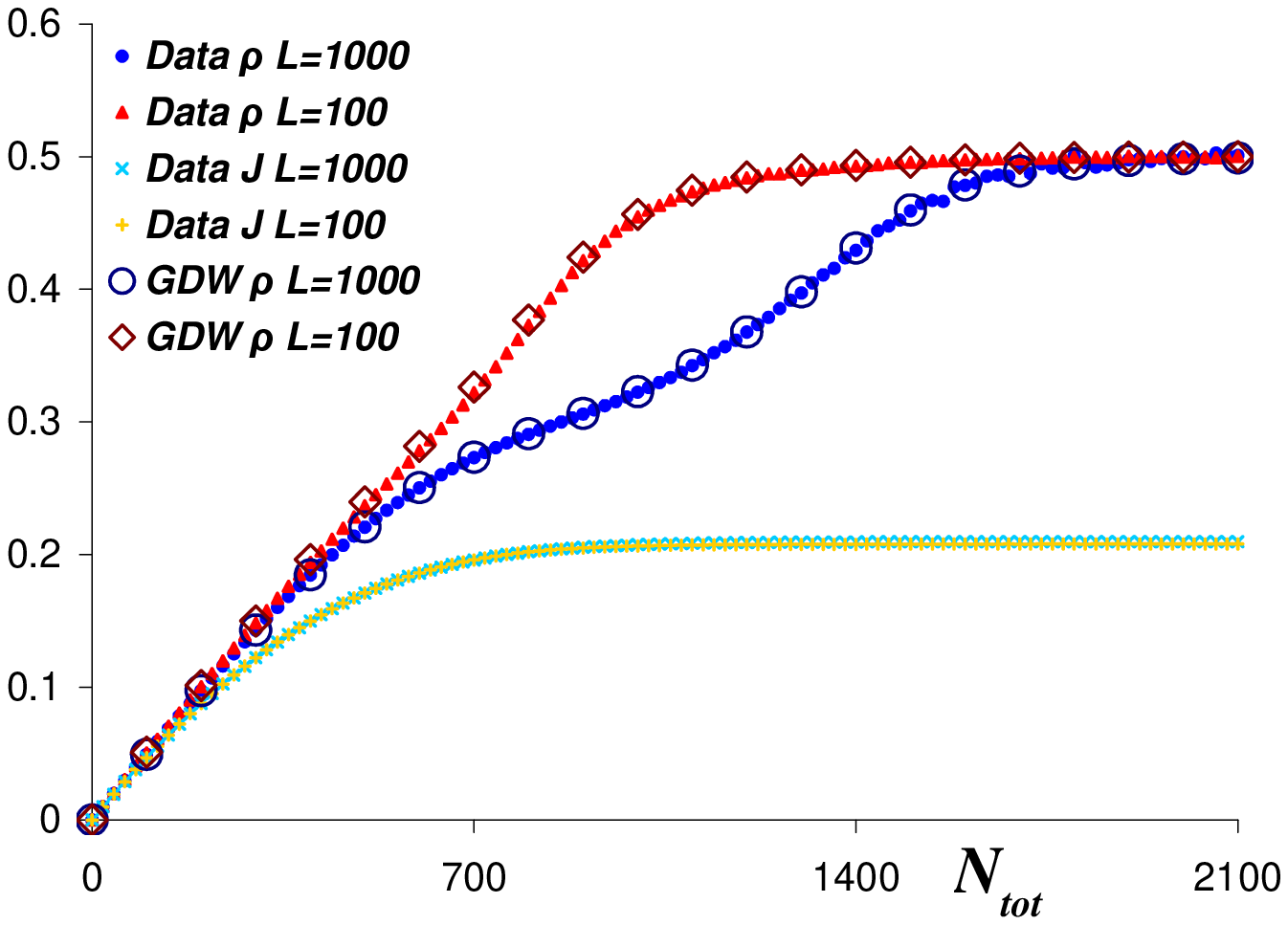} \label{SP-density-unequal} }
\caption{(Color online) Average overall density and current as a function of $N_{tot}$ for 
\subref{SP-density-equal} $L_1=L_2=1000$ and \subref{SP-density-unequal} $%
L_1=1000$ and $L_2=100$ with $\alpha =\beta =0.3$.\label{SP-density}}
\end{figure}

The SP case is the most challenging, due to the large fluctuations that
characterize this ``phase''. In the standard TASEP, the system exhibits a
freely moving shock, separating a low-density from a high-density region. In
the constrained TASEP, the crossover from the LD to the SP\ phase is highly
nontrivial. The details depend on both the length of the chain and the
functional form of $\alpha _{eff}$. Fortunately, the essentials are well
captured by DW theory \cite{DW,Santen02,ASZ08}, especially when it is
generalized appropriately \cite{CZ08}. Not surprisingly, two TASEPs of equal
length exhibit the same densities and currents as a function of $N_{tot}$.
Differences emerge only when $L_1\neq L_2$, as illustrated in Fig. \ref
{SP-density}, where $\alpha =\beta =0.3$ and $L_1=1000,$ $L_2=100$. These differences
may be expected, however, if we recall that single TASEPs with various
lengths behave quite differently when coupled to a finite pool \cite{ASZ08}.
Details of the origins of these differences in our case of two competing
TASEPs can be understood in terms of the generalized DW theory \cite{CZ08}
which we discuss in the next Section. Again, like the cases above, when finite size effects are taken into account, there are no qualitatively new phenomena when an additional chains is included in the competition for a finite pool of particles. 

\subsection{HD phase}

\label{HD-phase}

The most interesting results are observed with $\alpha $ and $\beta $ set in
the HD phase. For a single constrained TASEP, there are three distinct
regimes in the average total density \cite{ASZ08}, as $N_{tot}$ is varied.
The behaviors for small and large $N_{tot}$ are expected:\ For the former,
they follow the LD-behavior discussed above and for the latter, the density
and current approach their asymptotic values. The presence of the third
regime, interpolating between these two limits, was a surprise initially.
Here, the density rises linearly with $N_{tot}$ while the current remains
constant at its asymptotic limit. The origin of this linear dependence can
be traced to $N_p$ (the reservoir occupation) being essentially constant
over a range of $N_{tot}$, so that any changes in $N_{tot}$ are absorbed by
the lattice. In particular, $N_p$ takes the critical value $N_p^c$, given by 
$\alpha _{eff}\left( N_p^c\right) =\beta $, so that coexistence of low and
high density regions on the lattice is possible. Unlike the ordinary TASEP
however, the shock does not wander over the entire lattice. Instead, it is
localized at some point determined by $N_{tot}$, with an intrinsic width
controlled by $\partial _{N_p}f$ \cite{CZ08}. As for the current, it has
already reached $\beta (1-\beta )$ at the lower extreme of this linear
regime and so, it is not sensitive to the final crossover into the
asymptotic\ HD regime.

\begin{figure}
\includegraphics[width=8.2cm]{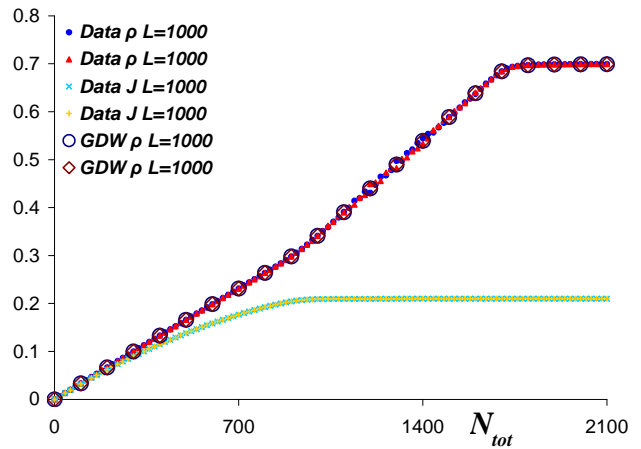}
\caption{(Color online) Average overall density and current as a function of $N_{tot}$ for $%
L_1=L_2=1000 $ with $\alpha =0.7$ and $\beta =0.3$.\label{HD-density-equal}}
\end{figure}

\begin{figure}
\includegraphics[width=8.2cm]{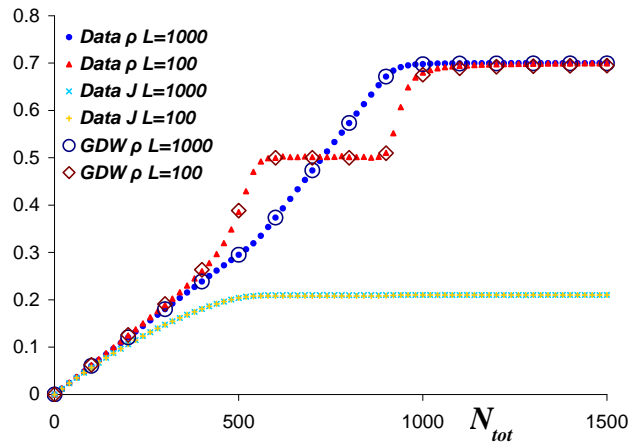}
\caption{(Color online) Average overall density and current as a function of $N_{tot}$ for $L_1=1000$
and $L_2=100$ with $\alpha =0.7$ and $\beta =0.3$.\label{HD-density-unequal}}
\end{figure}

Turning to a system with two TASEPs, there appear to be no new surprises if
they are of equal lengths, as displayed in Fig. \ref{HD-density-equal} for $%
L_1=L_2=1000$. The densities and currents for both TASEPs coincide and
exhibit the three distinct regimes found in the single TASEP case. In stark
contrast, however, a new feature emerges if $L_1\gg L_2$.\ To be specific,
we will discuss the case $L_1=1000,$ $L_2=100$ here, unless otherwise
stated. While the density of the longer TASEP still displays the expected
three regimes, the shorter system now develops \emph{five }regimes, as
illustrated in Fig. \ref{HD-density-unequal}. The two outermost regimes (for 
$N_{tot}$ near zero and $N_{tot}\rightarrow \infty $) are the familiar LD
and asymptotic regime. In the central region, however, we see a \emph{%
horizontal plateau}, bordered by two ``crossover'' regimes where the density
increases steeply with $N_{tot}$. The currents show no discernible signature
whatsoever, exhibiting only the LD and the saturation regime. To appreciate
the different behaviors in intuitive terms, we will present several
perspectives here. In the next section, we will provide the mathematical
approach, in which an exact solution of the generalized DW theory appears to
account for all data quite well.%

First, let us examine the ``plateau'' region. Similar to the case of a
single constrained TASEP, this region is characterized by $N_p$ being fixed
at $N_p^c$, and so, $\alpha _{eff}=\beta $ while changes of $N_{tot}$ are
absorbed by the \emph{two} TASEPs. The main difference here is that the
excess particles are free to choose either chain. Given that the pool
suffers only minor fluctuations, $N_1+N_2$ is also relatively constant, so
that the two chains simply trade particles back and forth. In terms of the
domain wall picture, a DW is present in each lattice, but their walks are
completely anticorrelated. Though the shocks are no longer localized as
before \cite{CZ08} and free to wander about, they are limited by $\left(
L_1,L_2\right) $. Specifically, the \emph{shorter }TASEP imposes the extent
by which the DW may wander on the longer lattice. In the ``plateau'' region, 
the DW is free to be anywhere on the shorter lattice, so that the average 
profile there is strictly linear. The overall density here is strictly $1/2$ and the ``plateau'' emerges. 
In this region, all changes of $N_{tot}$ are absorbed entirely by
the longer TASEP (i.e., $\partial \rho _1/\partial N_{tot}$ is precisely 
$1/L_1$ in Fig. \ref{HD-density-unequal}). However, as will be shown below,
the similarity of this behavior to the single TASEP case is deceptive. The
density profiles are quite distinct, reflecting the different origins of the
constraints on the DW walks. This picture also shows clearly why such a
``plateau'' cannot exist for the $L_1=L_2=L$ case: Neither TASEP imposes a
limit on the other. By symmetry, the only point when a DW on one lattice can
wander over one entire lattice is also the point where its counterpart can
wander over all of the other lattice. At this point, both profiles must be
linear and both densities are $1/2$, so that $N_1+N_2=L$. Since the pool
occupation must remain at $N_p^c$, this single point corresponds necessarily
to $N_{tot}=N_p^c+L$.

\begin{figure}
\subfigure[]{\includegraphics[width=8.2cm]{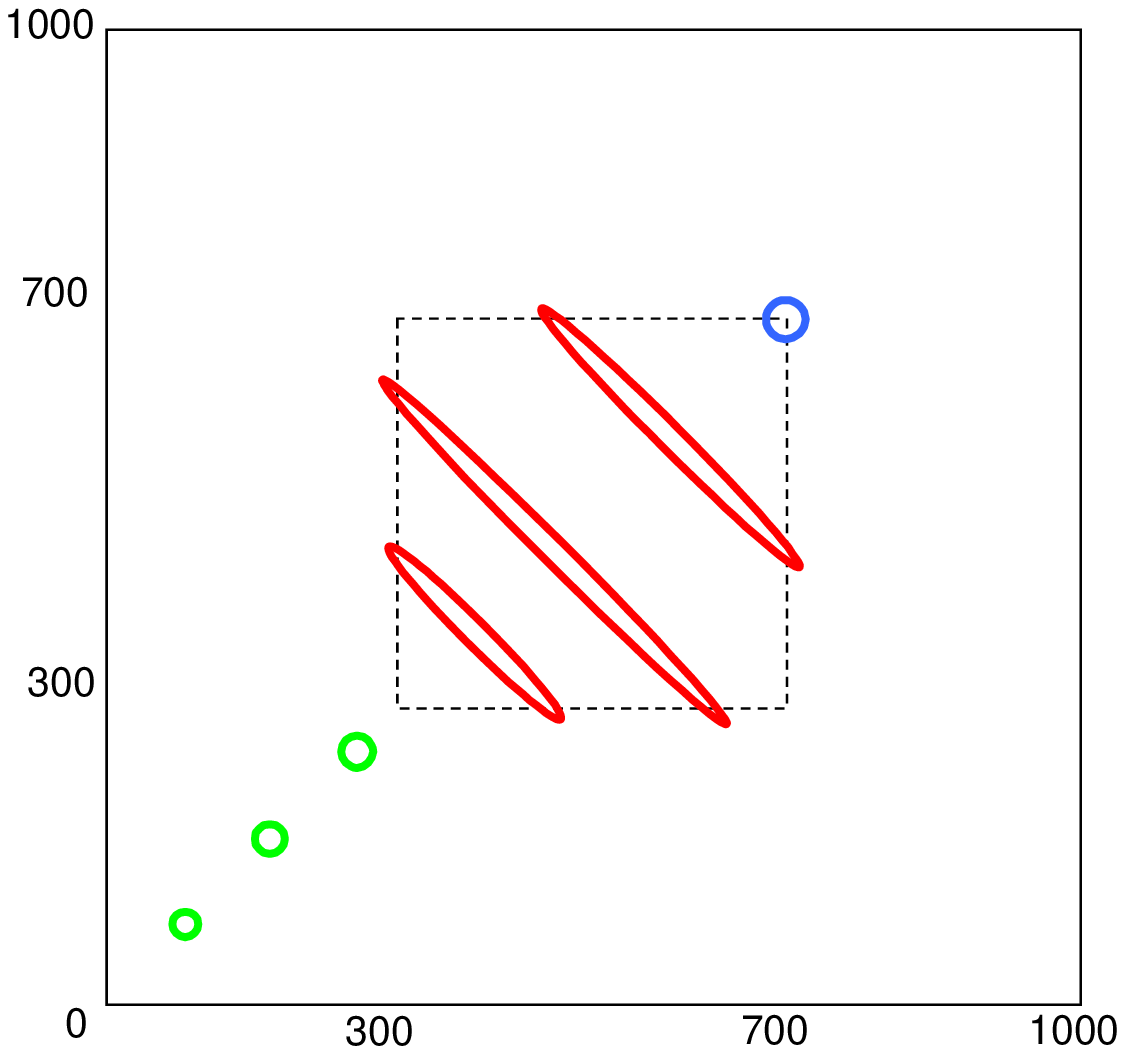} \label{sketch-equal}}
\subfigure[]{\includegraphics[width=8.2cm]{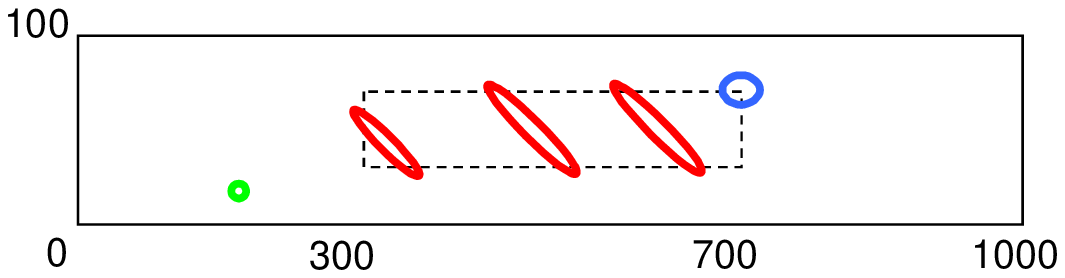} \label{sketch-unequal}}
\caption{(Color online) Rough sketches of distributions in the plane for \subref{sketch-equal} 
$L_1=L_2=1000$ and \subref{sketch-unequal} $L_1=1000$ and $L_2=100$. The dashed boxes 
correspond to the region allowed by LD/HD co-existence 
(i.e., densities of $0.3$/$0.7$ here).\label{sketch}}
\end{figure}

\begin{figure}
\includegraphics[width=8.2cm]{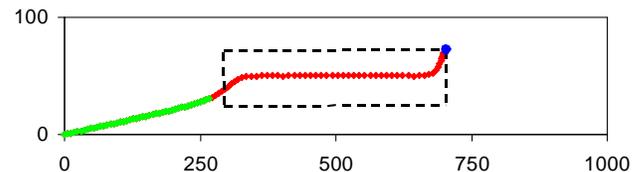}
\caption{(Color online) Simulation data for $L_1=1000$, and $L_2=100$ with $\alpha =0.7$ and 
$\beta =0.3$ in the $N_1$-$N_2$ plane. The dashed boxes correspond to the region 
allowed by LD/HD co-existence.\label{sketch-MC}}
\end{figure}

\begin{figure}
\subfigure[]{\includegraphics[width=8.2cm]{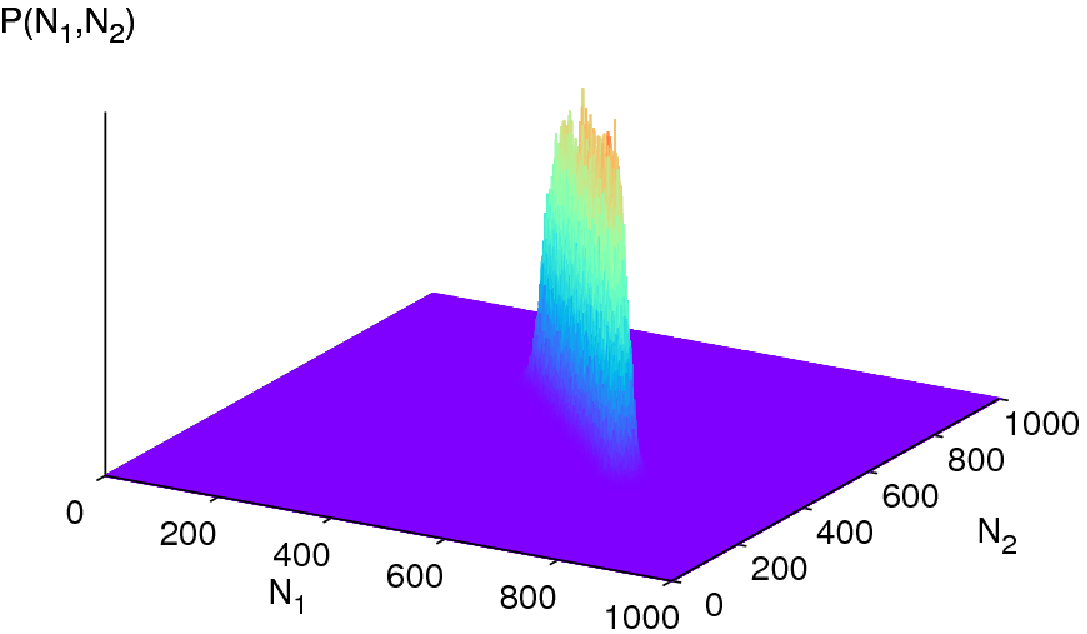} \label{HD-dist-equal} }
\subfigure[]{\includegraphics[width=8.2cm]{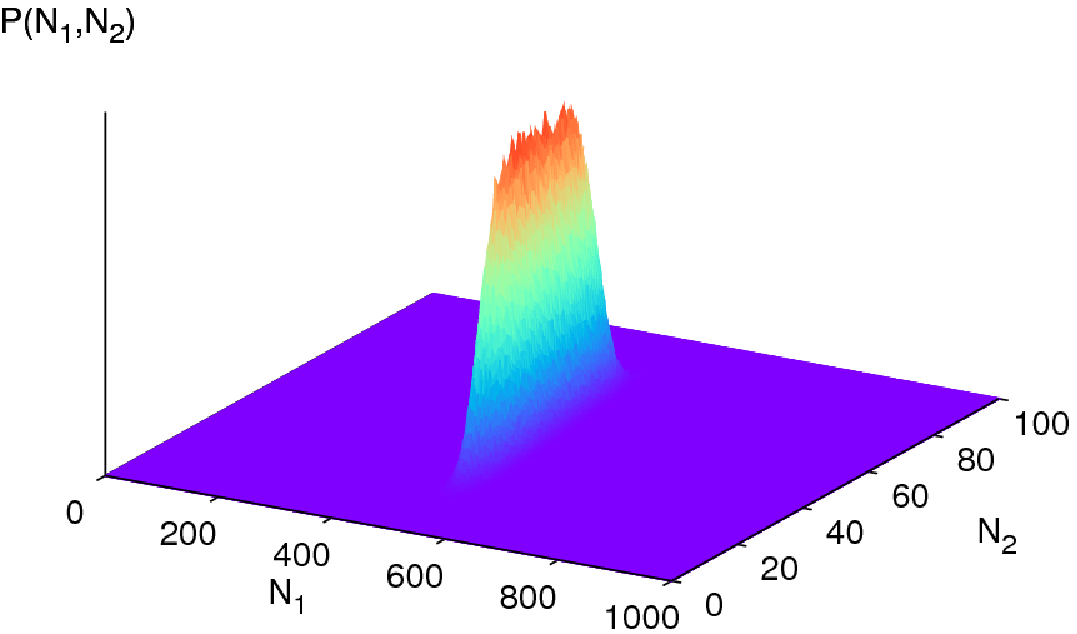} \label{HD-dist-unequal} }
\caption{(Color online) Distribution of $N_1$ and $N_2$ for \subref{HD-dist-equal} $N_{tot}=1400$ and 
$L_1=L_2=1000$ and \subref{HD-dist-unequal} $N_{tot}=700$, $L_1=1000$, and $L_2=100$ 
with $\alpha =0.7$, and $\beta =0.3$.  The z-axis scale for $P(N_1,N_2)$ is arbitrary in both figures.\label{HD-dist}}
\end{figure}

Another perspective on the existence of the ``plateau'' is offered in Fig. 
\ref{sketch}, which shows schematic views of the domains in the $N_1$-$N_2$
plane in which we are likely to find the $L_1=L_2=1000$ system (Fig. \ref{sketch}a) and the $L_1=1000,$ $L_2=100$ case (Fig. \ref{sketch}b). For very small $N_{tot}$, both
TASEPs will be in the LD phase, represented by the circles (green online)
near the origin. As $N_{tot}$ increases beyond this regime, the system will
find itself mostly in the long ellipses (red online), aligned along lines
of constant $N_1+N_2$. Finally, for large $N_{tot}$, both lattices will
saturate in the HD phase, indicated by the circle furthest from the origin
(blue online). These two simple sketches bring out the main difference
between the two cases: If the lattices are of unequal length, there is a
range of $N_{tot}$ for which the size of the ellipse remains the same. In
this range, the average density on the shorter TASEP is about $1/2$ (i.e., $%
N_2\cong 100/2$ here) while the occupation on the longer lattice continues
to increase. When the data in Fig. \ref{HD-density-unequal} are replotted in
the $N_1$-$N_2$ plane, this picture is largely confirmed: Fig. \ref
{sketch-MC}. In contrast, for a system with $L_1=L_2$, there is no such
comparable range and so, there is no ``plateau'' region. In Fig. \ref
{HD-dist}, we provide two examples of Monte Carlo data associated with these
sketches. They are histograms for finding $N_1$ and $N_2$ particles in the
two TASEPs. The ``ridges'' in these plots correspond to two of the ellipses
in Fig. \ref{sketch}: (a) along $N_1+N_2\cong 1100$ for the $L_1=L_2=1000$
case and (b) along $N_1+N_2\cong 525$ for the other case. These ridges also indicate that,
subjected to $N_1+N_2\cong const.$, all $(N_1,N_2)$ pairs are equally
likely. Of course, these observations reflect just the picture presented
above: the DWs on each chain are free to wander, but strongly anticorrelated
and limited by $(L_1,L_2)$. 

\begin{figure}
\includegraphics[width=8.2cm]{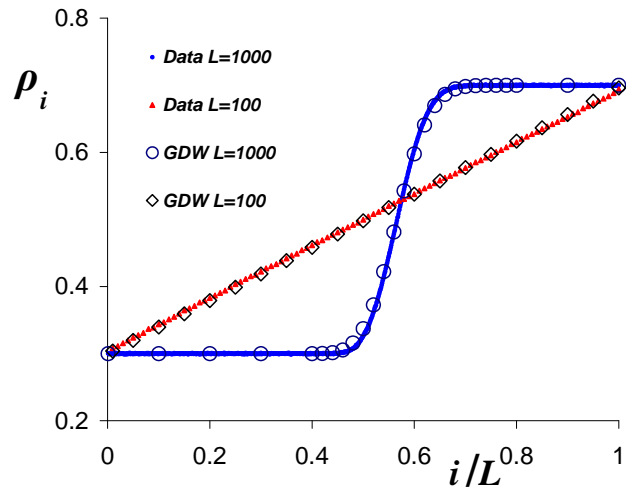}
\caption{(Color online) Average density profile for $N_{tot}=700$, $L_1=1000$, and $L_2=100$
with $\alpha =0.7$ and $\beta =0.3$.\label{HD-profile}}
\end{figure}

\begin{figure}
\includegraphics[width=8.2cm]{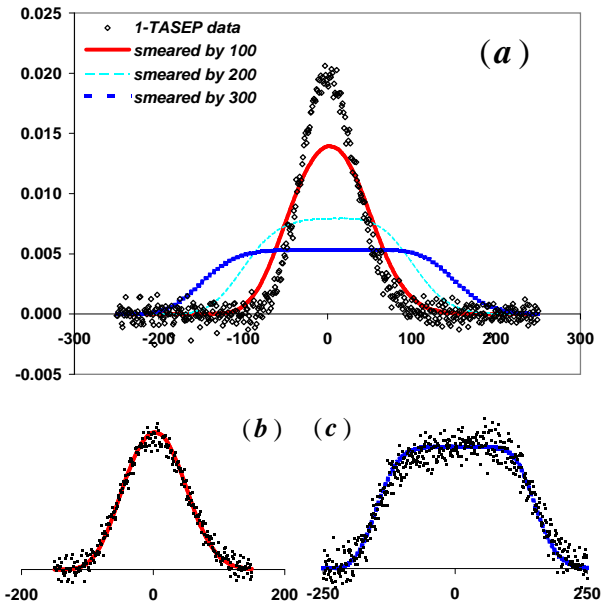}
\caption{(Color online) (a) Local slopes of the profile for a single TASEP with $L=1000$
and its ``smeared'' versions, shown here as lines (solid red, thin dashed green, short dashed blue). Open diamonds ($\lozenge$) are data points, obtained with $\alpha
=0.7,\beta =0.3,N^{*}=385,N_{tot}=650$, and shifted so that the peak lies at
the origin. (b),(c) Solid squares ($\blacksquare$) are data points for the $L=1000$
TASEP competing with a shorter one: $L=100$ (b) and $300$ (c), respectively. Lines
are same as those in (a).\label{smear profiles}}
\end{figure}

To provide yet another perspective, we present the density profiles of both
chains in the $L_1=1000,$ $L_2=100$ case in Fig. \ref{HD-profile}. Note that,
in this figure, the horizontal axis is the fractional distance ($i/L$) along
a chain and so, points on the two curves correspond to different absolute
distances ($i$). Being linear, the profile of the short TASEP clearly
reflects a totally delocalized DW. For the long chain, though the profile
indicates a localized shock, this appearance is deceptive. The width (over
which the profile changes from LD to HD) is actually somewhat larger than
its counterpart in a single constrained TASEP (see, e.g., Fig. 2 in
Ref. \cite{CZ08}). The origin of this broadening can be traced
to the additional fluctuations allowed by the shorter chain (approximately $%
100$ lattice sites in this case). In particular, we re-examined a single
TASEP of 1000 sites at the same $\alpha$ and $\beta $ and obtained its profile. 
To accentuate the ``interface'' of the profile, we plot in Fig. \ref{smear profiles}(a) the
local \emph{slopes} of the profiles (i.e., $\Delta \rho \left( i\right)
\equiv \rho \left( i+1\right) -\rho \left( i\right) $, shown as open
diamonds). Note that these can be regarded roughly as the probability of
finding the shock at site $i$. Next, we construct a ``smeared'' data set 
$\Delta \tilde{\rho}\left( i|m\right) $ as follows: Shift the raw data by 
$j=1, ..., m$ sites and then average over these $m$ data sets \cite{Note-smear}. 
By this procedure, we hope to account for the extra degree of freedom which the shock 
experiences, thanks to the presence of the short TASEP. 
The lines (solid red, thin dashed green, short dashed blue) in Fig. \ref{smear profiles}%
(a) illustrate the result of smearing with $m=100,200,$ and $300$. Now we
return to the two TASEP system and, in Fig. \ref{smear profiles}(b,c),
display $\Delta \rho \left( i\right) $ (solid squares) for the $L_1,L_2=1000,100$ and $1000,300$ cases, respectively. Meanwhile, the lines
here (solid red and short dashed blue, respectively) are precisely those from the smeared
profiles in Fig. \ref{smear profiles}(a). The good quantitative agreement
between the data and these $\Delta \tilde{\rho}\left( i|m\right) $ is a
convincing confirmation of the picture we presented: By freely exchanging
particles between the two chains, the shorter TASEP provides extra room for
an otherwise ``localized'' DW to wander in the longer chain. Our
conclusion here is that there are \emph{two} contributors to the
localization of the DW in a TASEP competing for finite resources. They are
(i) the feedback \cite{CZ08} due to a nontrivial $\alpha _{eff}$, producing
an ``intrinsic'' localization length, and (ii) the constraint from the other
TASEP participating in the give-and-take of particles. Clearly, (ii) means
that the longer chain imposes no constraint on the shorter one, so that the
shock is completely delocalized, regardless of mechanism (i). By the same
token, the profile of the longer TASEP is typically somewhat more complex,
as both mechanisms play a role. Naturally, if the chains are of equal
length, delocalization prevails on both and the profiles will be linear (to
the extent allowed by $N_{tot}$), as our simulations confirm. Now, despite
the effects of competition, it is possible to isolate the role of mechanism
(i) and observe an ``intrinsic'' profile, as follows. For each measurement
of $\left\{ n_\ell \left( i\right) \right\} $, we use the totals $N_{1,2}$
to estimate the position of the DWs in each lattice: $k_\ell $. Then, we
average the \emph{shifted} occupations $\left\{ n_\ell \left( i-k_\ell
\right) \right\} $ to arrive at the ``intrinsic'' profile. The result is
statistically identical to the profile in the single TASEP case \cite{CZ08},
so that we are confident of the merits of the intuitive picture presented
here. Details of this procedure and the comparisons will be provided
elsewhere \cite{Cook10}.

\begin{figure}
\includegraphics[width=8.2cm]{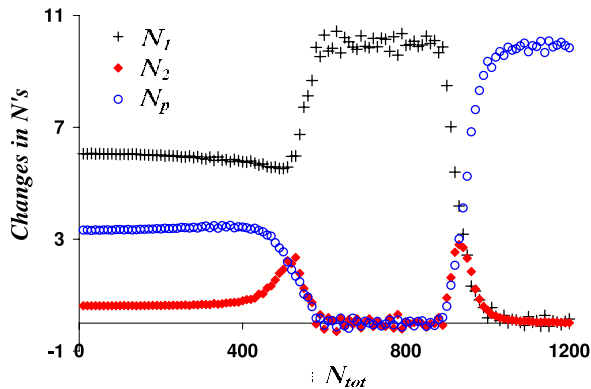}
\caption{(Color online) Occupation gradients of $N_1$, $N_2$, and $N_p$ with respect to $N_{tot}$ for 
$L_1=1000$, and $L_2=100$ with $\alpha =0.7$ and $\beta =0.3$.\label{Gradients}}
\end{figure}
Having addressed the central ``plateau'' regime, we turn to the two
bordering, ``crossover'' regions. In Fig. \ref{HD-density-unequal}, we see
that the behaviors displayed here are quite rich. To highlight these better,
we plot the \emph{gradients} of the three occupations, $N_{1,2,p}$,
associated with increasing $N_{tot}$ by $10$ particles. In Fig. \ref
{Gradients}, we show the data from the more interesting $L_1=1000,$ $L_2=100$
case. The regions corresponding to the central plateau and its borders are
most clearly seen for the small chain ($N_2$ , solid diamonds, red online):
two peaks with a flat valley in between. At a very naive level, these
features can be roughly understood from the sketch in Fig. \ref{sketch-unequal}.
For small $N_{tot}$, the circles (green online) traverse along a line with
slope $1/10$, corresponding to equal changes in the \emph{densities} of the
two chains. As $N_{tot}$ increases further, we see an ellipse (small, red
online) moving into the rectangle. Here, we might expect the two $N$'s to
increase together, until the ellipse spans the vertical range of the
rectangle. From there on, $N_2$ ceases to change while $N_1$ continues to
increase. A similar crossover region is present for the right end of the
rectangle, when $N_2$ again increases. Though this intuitive reasoning
provides a qualitatively picture of the five regimes, it clearly fails to
capture the details of the two crossover regions. For example, the changes
in the two $N$'s in the first crossover are \emph{anti}correlated, rather
than increasing together. Evidently, these details are sufficiently subtle
that they can only be fully understood in a quantitative theory for domain
wall motion -- the subject of Section \ref{sect4}.

\subsection{Competition between three TASEPs}

\begin{figure}
\includegraphics[width=8.2cm]{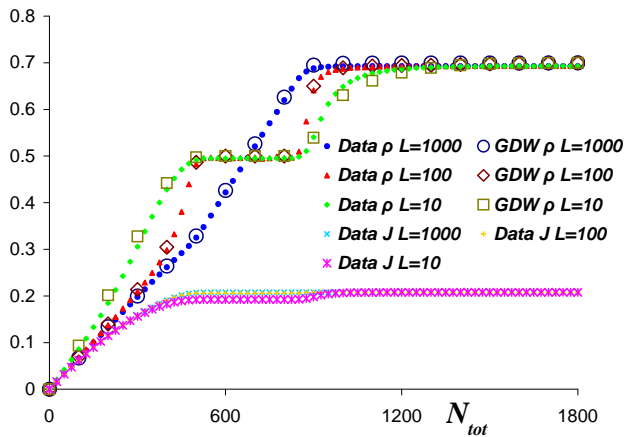}
\caption{(Color online) Average overall density and current as a function of $N_{tot}$ for $L_1=1000$, $%
L_2=100$, and $L_3=10$ with $\alpha =0.7$ and $\beta =0.3$.\label{HD-density-three}}
\end{figure}

\begin{figure}
\includegraphics[width=8.2cm]{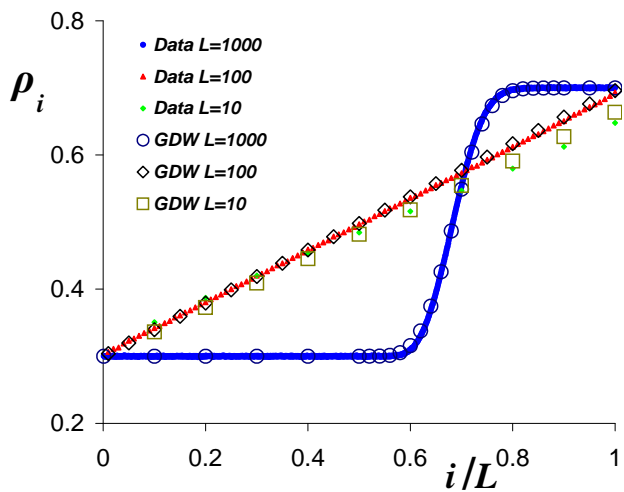}
\caption{(Color online) Average density profile for $N_{tot}=600$, $L_1=1000$, $L_2=100$, $%
L_1=10$ with $\alpha =0.7$ and $\beta =0.3$.\label{HD-profile-three}}
\end{figure}

Turning next to the study of competition between three TASEPs, we see
immediately that even more scenarios are possible, from all lengths being
equal, to some being the same, to all lengths being drastically different.
Though we have explored quite a few cases, we will only present data for the
most extreme one (HD): $L_1=1000,$ $L_2=100,$ $L_3=10$ with $\alpha =0.7$
and $\beta =0.3$. How the three densities vary as we increase $N_{tot}$ is
shown in Fig. \ref{HD-density-three}. We again see the smaller two TASEPs
entering into a ``plateau'' regime, with linear profiles and half filling on
the average. Meanwhile, the longest chain again displays a localized shock,
as shown in Fig. \ref{HD-profile-three}. In more detail, the shortest TASEP
reaches $\rho = 1/2$ first, followed by the intermediate chain. The behavior
of the longest TASEP is much the same as what we observed in the $L=1000$
chain above. It appears that adding a third chain does not lead to any
qualitatively novel behavior. The overall currents, especially if the severe finite size effects associated
with $L = 10$ are taken in account, also display few surprises. 
Of course, the phase space for $M=3$ is considerably larger and new
phenomena may very well emerge upon closer examination. 
Further studies are in progress and will be reported elsewhere. 
In the remainder of this paper, we will focus on
a quantitatively viable, analytic picture for an arbitrary number of TASEPs.

\section{Generalized domain wall theory}

\label{sect4}

To understand most of the phenomena we observed, it is sufficient to use the
simplest approximation \cite{ASZ08}, based on self-consistent equations
between the feedback dependence, $\alpha _{eff}\left( N_p\right) $, and the
occupation variables, $N_\ell $, for the TASEPs. The only serious
complication arises in the HD case, when $\alpha _{eff}$ reaches $\beta $
and each individual TASEP enters an SP-like regime. As presented above, a
variety of interesting behaviors emerge for which intuitively reasonable, simple
arguments paint a good qualitative picture. In this section, we will provide
a quantitative description, which relies on an excellent approximation,
namely, an appropriately generalized domain wall theory \cite{ASZ08,CZ08}.
Proposed about a decade ago for the standard single TASEP \cite{DW,Santen02},
DW theory assumes that the configurations are well accounted for by those with
a microscopic interface between two regions, one with high density $\rho _{+}$ 
and another with low density $\rho _{-}$. The generalization to a single
TASEP constrained by finite resources \cite{ASZ08,CZ08} provided excellent
agreement with all aspects of simulation data. Referring the reader to \cite
{CZ08} for details, let us first present a brief summary of this approach
here, in the context of \emph{multiple} TASEPs competing for the same pool
of particles.

We assume that a configuration of the system of $M$ TASEPs can be well
approximated by specifying the position of the shocks (DWs) on each 
lattice, $k_\ell
\in \left[ 0,L_\ell \right] $, with $\ell =1,\dots ,M$. Since all chains are
subjected to the same entry and exit rates of particles, we will further
assume that the densities before (sites $i\leq k$) and after (sites $i>k$)
the wall on all chains are identical, i.e., $\rho _{-}=\alpha _{eff}$ and $%
\rho _{+}=1-\beta $ , respectively. Now, $\alpha _{eff}$ depends on the
numbers in the pool and so, on the occupation on the lattices, $N_\ell $.
Thus, to close the equations, we need the relationship between $N_\ell $ and 
$k_\ell $: 
\begin{eqnarray}
N_\ell &=&\rho _{-}k_\ell +\rho _{+}(L_\ell -k_\ell ) \\
&=&\left( \alpha _{eff}+\beta -1\right) k_\ell +\left( 1-\beta \right) L_\ell
\label{N-k}
\end{eqnarray}
which leads us to the pool occupation: 
\begin{equation}
N_p=N_{tot}-\sum_\ell N_\ell
\end{equation}
and the dependence of $\alpha _{eff}$ on the $k$'s. In short, from $\alpha
_{eff}=\alpha f(N_p)=\alpha \tanh \left( \left[ N_{tot}-\sum_\ell N_\ell
\right] N^{*}\right) $, we have 
\begin{widetext}
\begin{equation}
\alpha _{eff}\left( K\right) =\alpha \tanh \left( \frac{N_{tot}-(1-\beta
)\sum_\ell L_\ell -\left( \alpha _{eff}+\beta -1\right) \sum_\ell k_\ell }{N*%
}\right)  \label{alpha-kl}
\end{equation}
\end{widetext}
Note that $\alpha _{eff}$ depends only on the sum 
\begin{equation}
K\equiv \sum_{\ell =1}^Mk_\ell 
\end{equation}
rather than the individual $k$'s. This simplification will be crucial for us
to find an exact steady state solution to the master equation in our system.
Of course, we must solve the non-linear, self-consistent equation (\ref
{alpha-kl}) to determine $\alpha _{eff}\left( K\right) $. This task was
performed numerically, as in the previous study \cite{CZ08}. Indeed, we
have the same functional form as before, except that we now encounter 
$\alpha _{eff}\left(K;\sum_\ell L_\ell \right) $ 
instead of $\alpha _{eff}\left( k;L\right) $.

Once $\alpha _{eff}\left( K\right) $ is known, the rates for the DW to hop
to the left ($D^{-}$) and right ($D^{+}$) can be computed. We denote
explicitly their $K$-dependence by a subscript: 
\begin{eqnarray}
D_K^{-} &=&\frac{\alpha _{eff}(1-\alpha _{eff})}{1-\beta -\alpha _{eff}} \\
D_K^{+} &=&\frac{\beta (1-\beta )}{1-\beta -\alpha _{eff}}
\end{eqnarray}
With these, the master equation for $P(\left\{ k_\ell \right\} ,t)$, the
probability to find DWs at $\left\{ k_\ell \right\} $ at time $t$, is well
defined. For $\left\{ k_\ell \right\} $ lying in the interior of the allowed
domain, it is easy to write 
\begin{eqnarray}
\partial _tP\left( \left\{ k_\ell \right\} ,t\right) &=&\sum_{j=1}^M\bigl[
D_{K+1}^{-}P\left( \left\{ k_\ell +\delta _{\ell j}\right\} ,t\right) \\
&&+D_{K-1}^{+}P\left( \left\{ k_\ell -\delta _{\ell j}\right\} ,t\right)\bigr] \nonumber \\
&&-M\left[D_K^{-}+D_K^{+}\right]P(\left\{ k_\ell \right\} ,t)  \nonumber \label{Full ME}
\end{eqnarray}
where $\delta _{\ell j}$ is the Kronecker delta. At the boundaries, we must
impose reflecting boundary conditions on the appropriate $k$'s. Since there
are quite a few ($3^M-1$) of these conditions, it will be helpful to begin 
by studying the $M=2$ case explicitly (4 corners and 4 sides of a rectangle). In the
next subsection, we will provide some details for obtaining the steady-state
solution associated with this case. We will find that the natural variables are 
$\left( K,Q\equiv k_1-k_2\right) $ rather than $\left( k_1,k_2\right) $. The
insights gained here will facilitate the analysis of the arbitrary $M$ case, 
to be presented in the last subsection.

\subsection{Case with two TASEPs}

Here, we focus on the steady-state
solution to Eqn. (\ref{Full ME}) for $M=2$, for $0<k_1<L_1$ and $0<k_2<L_2$.
Dropping the $t$, and setting the left to zero, this equation reduces to 
\begin{eqnarray}
0 &=&D_{K+1}^{-}[P(k_1+1,k_2)+P(k_1,k_2+1)]  \nonumber \\
&&+D_{K-1}^{+}[P(k_1-1,k_2)+P(k_1,k_2-1)]  \nonumber \\
&&-2[D_K^{-}+D_K^{+}]P(k_1,k_2)  \label{SS-in}
\end{eqnarray}
where 
\begin{equation}
K=k_1+k_2  \label{K2}
\end{equation}
here. The minimum boundary conditions to be imposed correspond to one or
both DWs being reflected from the ends of the lattices. Thus, at the four
sides of the rectangle, we write 
\begin{eqnarray}
0 &=&D_{k_1+1}^{-}[P(k_1+1,0)+P(k_1,1)] \\
&&+D_{k_1-1}^{+}P(k_1-1,0)-[D_{k_1}^{-}+2D_{k_1}^{+}]P(k_1,0)  \nonumber \\
0 &=&D_{k_2+1}^{-}[P(0,k_2+1)+P(1,k_2)] \\
&&+D_{k_2-1}^{+}P(0,k_2-1)-[D_{k_2}^{-}+2D_{k_2}^{+}]P(0,k_2)  \nonumber \\
0 &=&D_{L_2+k_1-1}^{+}[P(k_1-1,L_2)+P(k_1,L_2-1)] \\
&&+D_{L_2+k_1+1}^{-}P(k_1+1,L_2) \nonumber \\
&&-[D_{L_2+k_1}^{+}+2D_{L_2+k_1}^{-}]P(k_1,L_2) \nonumber \\
0 &=&D_{L_1+k_2-1}^{+}[P(L_1,k_2-1)+P(L_1-1,k_2)] \\
&&+D_{L_1+k_2+1}^{-}P(L_1,k_2+1) \nonumber \\
&&-[D_{L_1+k_2}^{+}+2D_{L_1+k_2}^{-}]P(L_1,k_2) \nonumber
\end{eqnarray}
Similarly, the conditions at the four corners are 
\begin{eqnarray}
0 &=&D_1^{-}[P(0,1)+P(1,0)]-2D_0^{+}P(0,0) \\
0 &=&D_{L_1+L_2-1}^{+}[P(L_1-1,L_2)+P(L_1,L_2-1)] \\
&&-2D_{L_1+L_2}^{-}P(L_1,L_2)  \nonumber \\
0 &=&D_{L_1-1}^{+}P(L_1-1,0)+D_{L_1+1}^{-}P(L_1,1) \\
&&-[D_{L_1}^{+}+D_{L_1}^{-}]P(L_1,0)  \nonumber \\
0 &=&D_{L_2-1}^{+}P(0,L_2-1)+D_{L_2+1}^{-}P(1,L_2) \\
&&-[D_{L_2}^{+}+D_{L_2}^{-}]P(0,L_2)  \nonumber
\end{eqnarray}
However, as discovered in a previous study \cite{CZ08}, there is a more
subtle boundary condition. For sufficiently low $N_{tot}$, it is not
possible for one or both DWs to reach the left boundary. Specifically, if $%
N_{tot}\leq $ $\left( 1-\beta \right) \left( L_1+L_2\right) $, then at least
one of the lattices cannot be filled with the high density, $\rho _{+}$, so
that the sum of the DW positions must be larger than $K_{\min }\equiv
L_1+L_2-N_{tot}/\left( 1-\beta \right) $. Another way to understand this
limit is that the pool is empty ($N_p=0$) when $K$ reaches $K_{\min }$. Both 
$\alpha _{eff}$ and $D^{-}$ vanish and we simply have $P\equiv 0$ for $%
k_1+k_2\leq $ $K_{\min }$. To summarize, we see that $P(k_1,k_2)$ can be
non-zero only in a (generally) ``cut-rectangular'' domain: 
\begin{equation}
\mathcal{R}:\quad 0\leq k_1\leq L_1,\quad 0\leq k_2\leq L_2,\quad K_{\min
}<k_1+k_2 
\end{equation}

Taking into account this complication of the boundary conditions, and thanks
to $D$ being dependent on only one variable, these equations can be solved
analytically. In particular, although the original competing TASEPs is a
non-equilibrium statistical mechanics problem, the DW approximation reduces
it to the point that detailed balance prevails. Specifically, the product of
the rates around every elementary loop (through the four points $\left(
k_1,k_2\right) ;\left( k_1+1,k_2\right) ;\left( k_1+1,k_2+1\right) ;\left(
k_1,k_2+1\right) $) in configuration space is 
\begin{equation}
D_K^{+}D_{K+1}^{+}D_{K+2}^{-}D_{K+1}^{-} 
\end{equation}
\emph{regardless} of the direction taken around the loop. Since all closed loops
in this space are composed of these elementary loops, the Kolmogorov
criterion \cite{KoCr} is satisfied. Thus, we have detailed balance 
\begin{equation}
D_K^{-}P(k_1,k_2)=D_{K-1}^{+}P(k_1-1,k_2)=D_{K-1}^{+}P(k_1,k_2-1)  \label{DB}
\end{equation}
and the solution can be obtained by recursion. Setting 
\begin{equation}
P(L_1,L_2)\equiv Z^{-1} 
\end{equation}
where $Z$ is a (normalization) constant, we easily find the stationary
distribution, for $K\leq L_1+L_2-1$, 
\begin{equation}
P(k_1,k_2)=Z^{-1}\Phi \left( K\right) \,\,,  \label{P*}
\end{equation}
\emph{provided} $(k_1,k_2)$ lies in $\mathcal{R}$. Here, we define 
\begin{equation}
\Phi \left( K\right) \equiv \prod_{j=K}^{L_1+L_2-1}\frac{D_{K+1}^{-}}{D_K^{+}%
}\,\,.  \label{Phi}
\end{equation}
and emphasize that $P(k_1,k_2)$ varies only via the sum $k_1+k_2$ and is
flat along lines of constant $K$. Due to the irregular shape of $\mathcal{R}$%
, the normalization factor $Z$ is straightforward to find but not simply $%
\sum_K\Phi \left( K\right) $, as in the single TASEP case \cite{CZ08}. For
example, if $0<K_{\min }<L_1$, then 
\begin{equation}
Z=1+\sum_{K=K_{\min }+1}^{L_1+L_2-1}\left[ \left( L_1-K\right) \Theta \left(
L_1-K\right) +L_2+1\right] \Phi \left( K\right)  \label{Z}
\end{equation}
where $\Theta $ is the Heavyside function.

Since this stationary distribution depends only on the sum of the shock
positions, it behooves us to change the variables from $(k_1,k_2)$ to $(K,Q)$%
, where 
\begin{equation}
Q\equiv k_1-k_2\,\,.
\end{equation}
In terms of these variables, Eqn. (\ref{SS-in}) becomes 
\begin{eqnarray}
0 &=&D_{K+1}^{-}[P(K+1,Q+1)+P(K+1,Q-1)] \\
&&+D_{K-1}^{+}[P(K-1,Q-1)+P(K-1,Q+1)]  \nonumber \\
&&-2[D_K^{+}+D_K^{-}]P(K,Q)  \nonumber
\end{eqnarray}
with similar replacements for the eight boundary conditions. The detailed
balance condition, Eqn. (\ref{DB}), now reads 
\begin{eqnarray}
D_{+K}^{+}P(K,Q)&=&D_{K+1}^{-}P(K+1,Q+1)\\
\nonumber&=&D_{K+1}^{-}P(K+1,Q-1)
\end{eqnarray}
and suggests a solution that depends only on $K$. Assuming the ansatz 
\begin{equation}
P(K,Q)=Z^{-1}\Phi \left( K\right) H\left( Q,K\right)  \label{P-Phi}
\end{equation}
where $H\left( Q,K\right) $ is a Heavyside-like function (unity in $\mathcal{%
R}$ and zero otherwise), we find $\Phi \left( K+1\right) \propto \Phi \left(K\right) D_K^{+}/D_{K+1}^{-}$. 
Taking some care with the boundary conditions, 
it is easy to verify that the form (\ref{P-Phi}) is indeed valid.

Before turning to a theory for the multi-TASEP case, let us compare the
predictions (with no adjustable parameters) of this approach with the data.
First, the density profiles can be obtained, following the methods in the
previous study \cite{CZ08}: 
\begin{eqnarray}
\rho _1(i) &=&\sum_{k_2=0}^{L_2}\Biggl[ (1-\beta)\sum_{k_1=0}^iP(k_1,k_2)\\
&&+\sum_{k_1=i+1}^{L_1}\alpha_{eff(K)}P(k_1,k_2)\Biggr] \nonumber \\
\rho _2(i) &=&\sum_{k_1=0}^{L_1}\Biggl[ (1-\beta)\sum_{k_2=0}^iP(k_1,k_2)\\
&&+\sum_{k_2=i+1}^{L_2}\alpha_{eff(K)}P(k_1,k_2)\Biggr] \nonumber
\end{eqnarray}
In this expression, it is clear that $\sum_{k_2}P(k_1,k_2)$ is just the
probability for finding the DW at $k_1$ regardless of the configuration in
the other TASEP and similarly for $\sum_{k_1}P$. We find that these agree
well with all observed profiles -- not only for regimes with little
structure, but also for complex situations such as the ``plateau'' region.
As an illustration, in Fig. \ref{HD-profile} we show profiles for the
familiar $L_1=1000,$ $L_2=100$ case with $N_{tot}=700$. It is clear that the
essentials of our system, such as a linear profile in one chain along with a
localized shock in the other, have been successfully captured in this
theory. From these profiles, we obtain the overall densities by $\rho _\ell
=\sum_i\rho _\ell (i)/L_\ell $. As shown in Figs. \ref{LD-density}, \ref
{SP-density}, \ref{HD-density-equal}, and \ref{HD-density-unequal}, they are
in excellent agreement with the simulation data. Not surprisingly, the
histograms for $\left( N_1,N_2\right) $ shown in Fig. \ref{HD-dist} can also be
predicted, being basically $P\left( k_1,k_2\right) $ via the correspondence 
(\ref{N-k}). Although the agreement is also quite good, we should remark on
two shortcomings. First, relationship (\ref{N-k}) between $N$ and $k$ cannot
be exact, since both $N$ and $k$ are integer-valued. Second, lattices with $%
N $ greater than $\rho _{+}L$ or less than $\rho _{-}L$ are obviously absent
from the theory, a limitation due to the DW approximation. However, in all
the regions we have explored, these shortcomings result only in minor
disparities.

Since so many aspects of our system can be understood by this approach, let
us return to form a better intuitive picture for the two ``crossover''
regions for the $L_1=1000,$ $L_2=100$ case discussed at the end of Section \ref
{HD-phase}. First, note that there are subtle changes in the gradients of $%
N_\ell $, as $N_{tot}$ increases up to the lower crossover ($%
N_{tot}\thicksim 400$, in Fig. \ref{Gradients}). By contrast, $\partial
N_p/\partial N_{tot}$ remains relatively constant. More significantly, the
longer chain ``loses'' while the shorter one ''wins.'' This behavior can be
traced to the DWs being mostly bound to the exit (right edge), but making
longer and longer excursions into the lattice as particles in the system
become more abundant. The exponential tails of this excursion are
essentially identical, provided they do not intrude significantly into the
entry (left edge). Indeed, if we measure the ratio of the profiles $\rho
_1\left( 900+i\right) /\rho _2\left( i\right) $, it is essentially unity for 
$N_{tot}$ up to $\sim 400$. Nonetheless, the overall density of the shorter
lattice is affected more by a similar portion of an enhanced profile (over $%
\rho _{-}$), and so, will increase faster. We should also remark that the
density to the left of the shock (low density region) in either lattice has
yet to reach the final value (i.e., $\rho _{-}<\beta $).

After we enter the first crossover region ($N_{tot}\sim 400$), the DW
wanders further from the exit in each lattice. It eventually reaches the left side of the smaller TASEP and enters the SP. To understand how this transition occurs, let us look at how the density of the ordinary,
unconstrained TASEP of infinite length changes as a function of 
$\alpha $
with $\beta <1/2$, as we cross the first order line $\alpha =\beta $.
Starting from $\alpha =0$, the density rises linearly until 
$\alpha =\beta $. At this point, the density jumps to $1/2$. It then jumps again to a value of $1-\beta $ for $\alpha >\beta $. For a system with a finite length, these jumps are no longer sharp, but smeared out near the value of $\alpha =\beta $, reflecting a rapid increase in the density (and number of particles) in this region. We now use this information, noting that (i) $N_{tot}$ controls the effective $\alpha $, and (ii) $\alpha _{eff}=\beta $ spans a whole region of 
$N_{tot}$, and (iii) that the shorter TASEP behaves essentially like an unconstrained one, since it can draw from both the pool and the longer chain. The first rapid increase, as $\alpha _{eff}$ approaches $\beta $ from below, is responsible for the first peak of $dN_{2}/dN_{tot}$ just above $N_{tot}=400$, shown in Fig. \ref{Gradients}. Clearly, the gradient then decays to zero as $N_{2}$ reaches the characteristic plateau. The next peak reflects the end of the $\alpha _{eff}=\beta $ region: it is related to the sharp increase of the density on the other side of the first order line. In terms of the domain wall picture, the first peak reflects the fact that the exponential tail in the density profile reaches the entrance and changes from an exponential decay into a linear one. In other words, the probability
of finding the shock near the entrance increases until it is flat across the whole system. At the second crossover, the reverse happens: the system is in a high-density phase, with a small exponential tail at the entrance, indicating that the shock is now predominantly found there.  

It is interesting to note that, in the first crossover regime, the pool
``loses'' steadily at sharing the increases in $N_{tot}$. Of course, over
the plateau regime, only the longer chain gains from the changes in $N_{tot}$. The picture for the second crossover regime is essentially the same, except occurring in ``reverse order.'' Needless to say, the pool is the ultimate ``winner'' in this competition, absorbing all increases in $N_{tot}$ beyond this regime.

\subsection{Multiple TASEPs}

The insights gained in the detailed analysis for two TASEPs greatly
facilitate investigation of the general multi-TASEP case here. In
particular, since the entry rates into all $M$ chains are the same, $\alpha
_{eff}$, and depend only on $K=\sum_\ell k_\ell $, the transition
probabilities in the master equation (\ref{Full ME}) again satisfy the
Kolmogorov criterion. A change of coordinates similar to the two TASEP case
can be performed, with $K\equiv \sum k_\ell $ as the special variable. The
boundary conditions are also straightforward though considerably more
tedious, since the $M$-dimensional generalization of $\mathcal{R}$ is more
complicated in general. Nevertheless, thanks to detailed balance, it is
simple to solve for the stationary distribution associated with this
(effectively one-dimensional) master equation. The answer will again be the
form (\ref{P-Phi}): 
\begin{equation}
P(\left\{ k_\ell \right\} )=Z^{-1}\Phi \left( K\right) H(\left\{ k_\ell
\right\} )
\end{equation}
where the Heavyside-like function is now defined for the $M$-dimensional $%
\mathcal{R}$. Needless to say, the ``cuts'' of constant $K$ across $\mathcal{%
R}$ are geometric objects of dimension $M-1$, the general shapes of which
are far more complicated than the lines in the $M=2$ case above. As a
result, the normalization factor is given by a much more complex expression
than (\ref{Z}), since the coefficients involve polynomials (in $K$) up to
order $M-1$. Nevertheless, these computations are very simple for modern
computers and $P(\left\{ k_\ell \right\} )$ can be easily accessed
numerically, for a reasonable range of $M$s. For example, we computed $%
P(k_1,k_2,k_3)$ for the three TASEP case discussed in the previous section ($%
L_1=10L_2=100L_3=1000$, with $\alpha =0.7=1-\beta $). From $P$, we obtained
the average profiles and overall densities, as a function of $N_{tot}$. As
shown in Fig. \ref{HD-density-three}, there is excellent agreement between
this theory and the data, over the many regimes encountered while $N_{tot}$
increases. Similarly, we find good agreement for the average density
profiles, as illustrated in Fig. \ref{HD-profile-three} for the case with $%
N_{tot}=600$. While two profiles are linear and one contains a localized
shock, all properties are well predicted by this approach. Of course, the
crossover regimes are richer, as each of the shorter chains produces peaks
in the gradients similar to those shown in Fig. \ref{Gradients}. With no
qualitatively new phenomena, these and other cases, as well as further
details, will be discussed in a later publication \cite{Cook10}. Our
conclusion is that the generalized DW theory is remarkably successful at
capturing the essence of our problem. Only the study of very sensitive
quantities reveals a poorer match between this approach and simulation
data.

To end this section, let us note that there is a thermal analog for 
$P(\left\{ k_\ell \right\} )$, namely, the canonical ensemble. Here, $K$
plays the role of the total energy, $E$, while the total number of points in
a sheet of fixed $K$ corresponds to the microcanonical partition function, 
$\Omega \left( E\right) $. The point $\left\{ k_\ell \right\} =\left\{ L_\ell
\right\} $ (highest $K$ allowed) would be the ``ground state,'' while the
precise connection between $E$ and $K$ is given by $E=-k_BT\ln \Phi \left(
K\right) $. Of course, we chose to label our normalization constant $Z$ to
carry this analogy to its logical end. Meanwhile, $N_{tot}$ seems to play
the role of temperature, with the average of $-K$ being a monotonically
increasing function of $N_{tot}$. Exploring this correspondence will be both
interesting and imperative, especially if we hope to make progress toward
our goal, namely, a cell with a few thousand different genes, each appearing
in hundreds of copies.

\section{Summary and Outlook}

In this paper, we explored how competition between TASEPs affects the
density profile, the overall density, and the current for each chain. A
feedback mechanism introduced previously \cite{ASZ08} was implemented. We
used Monte Carlo simulations to explore the properties of the overall
densities, currents, and profiles of the TASEPs, for a variety of parameters
(lattice length, $N_{tot}$, $\alpha $, and $\beta $). The competition
produced several novel features that are absent from the case of a single
TASEP constrained by finite resources \cite{ASZ08,CZ08}. There, the feedback
serves to localize a domain wall, when the control parameters are set in
favor of its appearance in the lattice. Here, the presence of \emph{other}
TASEPs adds an extra dimension to the feedback, leading to the
delocalization of the DW to the extent allowed by the lengths of the
other chains. Thus, when a long chain competes with a short one, its DW
wandering is limited by the length of the latter. By contrast, a DW in a
short chain is free to roam over the entire lattice, so that the average
profile displayed is strictly linear and the average overall density is just 
$1/2$. This picture can be readily generalized to three or more chains and
is confirmed in limited simulation studies with three competing TASEPs.

For the single TASEP with feedback, the standard domain wall theory 
was appropriately
generalized and proved to be extremely successful \cite{CZ08}. Extending
this theory to an arbitrary number of TASEPs is straightforward and a master
equation for $P\left( \left\{ k_\ell \right\} ,t\right) $, where $k_\ell $
denotes the position of the DW in the $\ell ^{th}$ chain, is easily
formulated. Fortunately, we are able to find the steady-state solution in a
system where the entry rate onto all chains are the same. The details were
presented for the two TASEP case and steps for extending it to arbitrary $M$
were provided. A remarkable features is the existence of an intimate mapping
from our steady state to a canonical equilibrium ensemble. From this
stationary distribution, all density profiles and currents can be predicted,
with \emph{no} adjustable parameters. The results agree well with all data
and give us much insight into a variety of phenomena discovered first in
simulations.

The most intriguing questions beyond our study here focus on time-dependent
phenomena. Even for the standard TASEP, there is a wealth of interesting
dynamics \cite{RingDyn,OpenDyn}. How are these affected by constraints of
finite resources and competition? For example, one of the simplest
quantities displaying remarkable behavior in the open TASEP is the power
spectrum associated with the total occupation, $N\left( t\right) $ \cite
{AZS07}. Preliminary data for a single chain coupled to finite resources
reveal a host of novel phenomena \cite{CZ10}. With many chains competing for
one pool of particles, we can study many other quantities, such as
correlations between the various TASEPs. Hopefully, these pursuits will
reveal other exciting secrets in this system and provide us with a deeper
insight. 

Being motivated by protein synthesis in cells, there are many extensions we can explore. Here we list some examples.
The initiation rates for various genes are far from being identical. Thus, we should introduce a full set of $\left\{ \alpha _\ell \right\}$s, to model 
highly and rarely expressed genes. Naturally, these
will induce complex $\left\{ k_\ell \right\} $-dependences in the DW hopping
rates: $D^{\pm }\left( \left\{ k_\ell \right\} \right) $. One serious
consequence is that the transition rates in the master equation (\ref{Full
ME}) will typically violate the Kolmogorov condition, so that the 
steady-state solution will be truly non-equilibrium in character. Nontrivial
steady-state probability currents necessarily follow \cite{ZS07}, and their
implications surely deserve further pursuit. Other obvious extensions include
important aspects of protein synthesis which have been considered in earlier models, such as having particles with finite extent to model the fact that ribosomes are relatively large molecules ``covering'' many codons \cite{bio}, and the inclusion of inhomogeneous hopping rates along the lattice to model the inhomogeneous sequence of codons and the wide range of concentrations of their associated aa-tRNAs \cite{Dong08}. Another aspect is the ribosome recycling enhancement considered by T. Chou \cite{TChou03} and the role of diffusion of (the subunits of) ribosomes in a competitive environment. Along these lines, to model the workings of a cell better, we should consider a system with some regulation on $N_{tot}$, as opposed to being just a preassigned, fixed number. Finally, an even more ambitious goal is to include not only the competition for ribosomes, but also for the many
varieties of aa-tRNA molecules. Clearly, much work remains to be done in order
to arrive at a realistic model of, and to better understand, protein
synthesis in a cell.

\begin{acknowledgments}
The authors would like to thank D.A. Adams, T. Chou, B. Derrida, Jiajia Dong, E. Levine, and G.M. Sch\"{u}tz for stimulating discussions. This work is supported in part by the US National Science Foundation through DMR-0705152.
\end{acknowledgments}


\begin{thebibliography}{99}
\bibitem{Note-smear}  Alternatively, $\Delta \tilde{\rho}\left( i|m\right) $
can be regarded as a coarse-grained average of $\Delta \rho \left( i\right) $%
, given by \\
$\sum_{j=0}^m \Delta \rho \left( i+j-m/2\right) / \left(m+1\right) $, \\
or simply \\$\left[ \rho \left( i+1+m/2\right) - \rho \left( i-m/2\right) \right] 
/ \left(m+1\right) $ .
\end{thebibliography}
\end{document}